\documentclass{amsart}%
\usepackage{amsfonts}
\usepackage{amsmath}
\usepackage{amssymb}
\usepackage{graphicx}%
\theoremstyle{plain}
\newtheorem{theorem}{Theorem}[section]

\newtheorem{corollary}{Corollary}

\newtheorem{definition}{Definition}

\newtheorem{lemma}[theorem]{Lemma}

\newtheorem{proposition}{Proposition}

\numberwithin{equation}{section}
\begin{document}
\title[A new Proof of semicircle law of fixed trace square ensemble]{A new Proof of semicircle law of fixed trace square ensemble}
\author[Da Xu and Lihe Wang]{Da Xu and Lihe Wang}
\address{Department of Mathematics\\
The University of Iowa\\
Iowa City, IA 52242} \email{dxu@math.uiowa.edu,lwang@math.uiowa.edu}

 \begin{abstract}
    In the present paper, we give a simple proof of the level density of fixed trace square ensemble.
We derive the integral equation of the level density of fixed trace square ensemble.Then we analyze the asymptotic
behavior of the level density. 
\end{abstract}

 \maketitle

\section{Introduction\label{Intro}}
E.P.Wigner in \cite{W1} and \cite{W2} showed that the level density
(one point correlation) of "border matrix" ensemble which is a kind
of real symmetric matrix ensemble, asymptotically approaches to the
semicircle,i.e.,
\begin{equation}
\sigma_{SYS,N}(x)\rightarrow \sqrt{2N-x^2}/\pi.
\end{equation}
Porter and Rosenzweig also found that the level density of some
special random matrices satisfy semi-circle law
\cite{PR1}\cite{PR2}.

 Wigner's movitation arose from the consideration of properties of wave functions of quantum mechanical
 system  that are extremely complicated that statistical considerations can be applied to them. From a computational point of view,
random matrix theory, as a statistical theory, is very important to
describe extremely complicated quantum mechanical systems such as
large atoms, QCD physics(see \cite{T} for review). It is
instructive to continue to work on the properties of RMT in the
problems of complicated quantum systems.

    Recently there has been work on the analysis of top eigenvalues of random matrices(\cite{B}).When we evaluate some quantities, for example the ratio of
the expectation of the largest square of eigenvalues and the
expectation of the square of an arbitrary eigenvalues
\begin{equation}
\frac{\langle\max_{1\leq i\leq N}{x_i^2}\rangle_{GUE}}{\langle
x_1^2\rangle_{GUE}}=\frac{\int\cdots\int \max_{1\leq i\leq N} x_i^2
e^{-\sum_1^N x_i^2} \prod_{i<j}(x_i-x_j)^2
dx_1dx_2\cdots dx_N}{\int\cdots\int \max_{1\leq i\leq N} x_1^2 e^{-\sum_1^N
x_i^2} \prod_{i<j}(x_i-x_j)^2 dx_1dx_2\cdots dx_N}
\end{equation}, we can change the integrals into polar coordinates. We can see the integration on vandermonde space is the
difficulty. Therefore it is important to consider the "fixed trace square ensemble" on $\mathbb{R}^N$
\begin{equation}
P_{v,N}(x_1,x_2,\cdots,x_N)=C_v\prod_{i<j}|x_i-x_j|^2,
\end{equation}
on the $N-1$ dimensional unit sphere $\{(x_1,\cdots,x_N);\sum_1^N
x_i^2=1\}$ and where $C_v$ is a normalization constant.
Stieltjes  in 1914 proved a theorem that states the maximum value of vandermonde polynomial $\prod_{i,j}|x_i-x_j|^2$ in the ball
$\{x_1^2+x_2^2 \cdots +x_N^2\leq N(N-1)\}$ achieves its maximal values at zeros of Hermite polynomials. In \cite{B}, the author considered the bounded trace square ensemble and obtained semicircle law of bounded trace square ensemble.
In \cite{ACMA}, the authors reobtained the semicircle law of fixed trace square ensemble by comparing it with GUE and saddle point method. In the present paper, we would like to give a simple proof of the semicircle law of fixed square ensemble. 

\section{integrals of fixed trace square ensemble}
Let's give a general definition of the ensemble on a homogeneous
space.
\begin{definition}
 $P(X)=P(x_1,x_2,\cdots,x_N)$ is  a  probability distribution
satisfying  on a homogenous space $M$ with respect to the permutation
group action on $x_1,x_2,\cdots,x_N$, which are the local coordinates. The level density is defined to be
\begin{equation}
\sigma_{P}(x)=N\int P(x_1,x_2,\cdots,x_N) dx_2\cdots dx_N.
\end{equation}
\end{definition}
In this paper, we are concerned with the case that $M$ is the $N-1$
dimensional sphere.

We would like to compute the level density of the ensemble
$P(x_1,x_2,\cdots,x_N)=C_N \prod_{i<j}|x_i-x_j|^2 $ on the $N$
dimensional ball $B_1=\{(x_1,x_2,\cdots,x_N);\sum_{i=1}^N x_i^2<1\}$. The
level density on sphere will easily follow. We will use Selberg's
integral \cite{S}.
\begin{theorem}(Selberg) For any positive integer $n$, let $dx\equiv
dx_1\cdots dx_N$,

\begin{align}
\Delta(x)& \equiv\Delta(x_1,\cdots,x_2)=\prod_{1\leq j<l\leq
n}(x_j-x_l)
, \ if \ n >1,\nonumber\\
\Delta(x)&=1, \ if \ n=1,
 \end{align}
and
\begin{equation}
\Phi(x)\equiv \Phi(x_1,\cdots,x_N)=|\Delta(x)|^{2\gamma}\prod_{j=1}^N
x_j^{\alpha-1}(1-x_j)^{\beta-1}.
\end{equation}
Then
\begin{equation}
I(\alpha,\beta,\gamma,n)\equiv\int_0^1\cdots\int_0^1
\Phi(x)dx=\prod_{j=0}^{n-1}\frac{\Gamma(1+\gamma+j\gamma)\Gamma(\alpha_j\gamma)\Gamma(\beta+j\gamma}
{\Gamma(1+\gamma)\Gamma(\alpha+\beta+(n+j-1)\gamma},
\end{equation}
and for $1\leq m\leq n$,
\begin{equation}
\int_0^1\cdots\int_0^1 x_1 x_2\cdots x_m\Phi(x)dx=\prod_{j=1}^m
\frac{\alpha+(n-j)\gamma}{\alpha+\beta+(2n-j-1)\gamma}\int_0^1\cdots\int_0^1
\Phi(x)dx,
\end{equation}
valid for integer $n$ and complex $\alpha$,$\beta$,$\gamma$ with
\begin{equation}
Re\alpha>0,
Re\beta>0,Re\gamma>-\min(\frac{1}{n},\frac{Re\alpha}{n-1},\frac{Re\beta}{n-1}).
\end{equation}
\end{theorem}

We need to compute the constant in the expression of the ensemble.
We cannot directly apply Selberg's theorem, because the domain we
integrate on is the unit ball instead of $\mathbb{R}^N$. By the
virtue of gamma function, we can change the integral on the unit
ball to the integral on $\mathbb{R}^N$ of Selberg's type. Let us
prove the following identity first.
\begin{proposition}
For any $R>0$, we have
\begin{equation}
\int_{\sum_{i=1}^N x_i^2\leq R^2} \Delta^2 dx=
R^{n^2}\frac{1}{\Gamma{(n^2/2+1)}}(2\pi)^{n/2}2^{-N^2/2}
\prod_{j=1}^N \Gamma(1+j )
\end{equation}
\end{proposition}

\begin{proof}
In Selberg's integral, letting
$x_i=\frac{y_i}{2L}+\frac{1}{2}$\cite{M} and
$L\rightarrow\infty$,then
\begin{equation}
\int_{-\infty}^{\infty}\cdots\int_{-\infty}^{\infty} |\Delta|^{2\gamma}
\prod_{j=1}^N
e^{-ax_j^2}dx_j=(2\pi)^{n/2}(2a)^{-N(\gamma(n-1)+1)/2}\prod_{j=1}^N
\frac{\Gamma(1+j\gamma)}{\Gamma(1+\gamma)}\label{114}
\end{equation}
Multiplying $a^{\beta-1}e^{-a}$ to (\ref{114})and integrating with
respect to $a$, we get
\begin{align}
&\int_{-\infty}^{\infty}\cdots\int_{-\infty}^{\infty}
|\Delta|^{2\gamma} \prod_{j=1}^N (1+\sum_{i=1}^N
x_i^2)^{-\beta}dx_j\\
=&\frac{\Gamma(\beta-N(\gamma(n-1)+1)/2)}{\Gamma(\beta)}(2\pi)^{n/2}2^{-N(\gamma(n-1)+1)/2}
\prod_{j=1}^N \frac{\Gamma(1+j\gamma)}{\Gamma(1+\gamma)}.
\end{align}
We change variables by $x_i=\frac{y_i}{\sqrt{1-\sum_{j=1}^N
y_i^2}}$. The determinant is
\begin{equation}
\frac{r_x^{n-1}dr_x}{r_y^{n-1}dr_y}=\frac{1}{(1-\sum_{j=1}^N
y_j^2)^{\frac{n-1}{2}}}\frac{1}{(1-\sum_{j=1}^N
y_j^2)^{\frac{3}{2}}}=\frac{1}{(1-\sum_{j=1}^N
y_j^2)^{\frac{n}{2}+1}},
\end{equation}
where $r_x$,$r_y$ are the radius coordinates in the polar coordinate
system. Therefore we have
\begin{align}
&\int_{-\infty}^{\infty}\cdots\int_{-\infty}^{\infty}
|\Delta|^{2\gamma} (1+\sum_{i=1}^N
x_i^2)^{-\beta}\prod_{j=1}^N dx_j\\
=& \int_{\sum_{j=1}^N y_j^2 \leq 1}
|\Delta(y)|^{2\gamma}(1-\sum_{i=1}^N
y_i^2)^{\beta-N(\gamma(n-1)+1)/2)-1}\prod_{j=1}^N dy_j.\label{1.19}%
\end{align}
Then we end up with
\begin{align}
&\int_{\sum_{j=1}^N y_j^2 \leq 1}
|\Delta(y)|^{2\gamma}(1-\sum_{i=1}^N
y_i^2)^{\beta-N(\gamma(n-1)+1)/2)-1}\prod_{j=1}^N dy_j\nonumber\\
=&\frac{\Gamma(\beta-N(\gamma(n-1)+1)/2)}{\Gamma(\beta)}(2\pi)^{n/2}2^{-N(\gamma(n-1)+1)/2}
\prod_{j=1}^N \frac{\Gamma(1+j\gamma)}{\Gamma(1+\gamma)}.
\end{align}
Now set $\gamma=1$ and $ \beta-N(\gamma(n-1)+1)/2)-1=0$. We get
\begin{equation}
\int_{\sum_{j=1}^N y_j^2 \leq 1} |\Delta(y)|^{2}\prod_{j=1}^N dy_j
=\frac{1}{\Gamma(n^2/2+1)}(2\pi)^{n/2}2^{-N^2/2} \prod_{j=1}^N
\Gamma(1+j ).
\end{equation}\label{1.21}
By rescaling, we conclude the result of this proposition.
\end{proof}

we are now ready to prove the following identity.
\begin{theorem}
The level density of $P(x)=const.\Delta^2(x)$ in the unit ball is
\end{theorem}
\begin{proof}
The level density of $P(x)$ is
\begin{align}
\sigma_P(y_1)&=N\int\cdots\int_{y_2^2+y_3^2+\cdots+y_N^2\leq
1-y^2}dy_2\cdots dy_N
\Delta^2(y)\nonumber\\
&=N\int\cdots\int_{y_2^2+y_3^2+\cdots+y_N^2\leq 1-y^2}dy_2\cdots dy_N
\prod_{j=2}^N (y_1-y_j)^2 \prod_{2\leq i\leq j\leq N}(y_i-y_j)^2\nonumber\\
&=N\int\cdots\int_{y_2^2+y_3^2+\cdots+y_N^2\leq 1-y^2}dy_2\cdots dy_N
  \prod_{j=2}^N (y_1^2-2y_1y_j+y_j^2)\prod_{2\leq i\leq j\leq N}(y_i-y_j)^2\nonumber\\
&=N\int\cdots\int_{y_2^2+y_3^2+\cdots+y_N^2\leq 1-y^2}dy_2\cdots dy_N
   (\sum_{p=0}^{N-1}\sum_{q=0}^{N-1-p}y_1^p ((-2)^q y_1^q )())\prod_{2\leq i\leq j\leq N}(y_i-y_j)^2\nonumber\\
&=N(\sum_{p=0}^{N-1}\sum_{q=0}^{N-1-p}(-2)^q
y_1{p+q}(1-y_1^2)^{1/2(-1-2p-q+N^2}\langle y_2 y_3\cdots
y_{q+1}y_{q+2}^2 y_{q+3}^2\cdots y_{N-p}^2 \rangle_{P,B_1}
\end{align}
\end{proof}

One can easily apply Aomoto's technique (\cite{A}) to compute
$\langle x_1 x_2 \cdots x_m\rangle$. However,we wish that we could get
an analytic expression of the level density. Unfortunately, as wee
have seen the above theorem, we have to explicitly write
expectations such as $\langle x_1^2 x_2^2\cdots x_m^2
x_{m+1}x_{m+2}\cdots x_n\rangle$ which turns out to be very hard and
remains an open problem. This fact makes us have to think about the
numerical solution of the level density.

\section{the integral equation of level density}
We would like to derive the integral equation of the level density
of  fixed trace square ensemble in another way. The idea is this. Consider the
level density of GUE. On the sphere with radius $r$, the level
density is $\frac{1}{r}\Phi(\frac{x}{r})$ with weight
$C_{N}^{-1}e^{-r^2}r^{n-1}r^{n(n-1)}$ where
\begin{equation}
C_{N}=\int_0^\infty e^{-r^2}r^{N-1}r^{N(N-1)}
dr=\frac{1}{2}\Gamma(\frac{N^2}{2}).
\end{equation}

So we have
\begin{theorem}
The level density of fixed trace square ensemble satisfies the integral equation
\begin{equation}
2\Gamma^{-1}(\frac{N^2}{2})\int_x ^\infty e^{-r^2} r^{N^2-2}
  \sigma_{\Phi,n}(x/r)dr=\sigma_{GUE,n}(x).\label{124}%
\end{equation}
\end{theorem}

\begin{proof}
For any $f(x_1)\in C^\infty_0(\mathbb{R})$, we change the following
integral into polar coordinate sytem.
\begin{align}
&\int \sigma_{GUE}(x_1)f(x_1)dx_1 \nonumber\\
=&N \int C_{GUE,N}e^{-\sum_1^N x_i^2} \prod_{i<j}(x_i-x_j)^2 f(x_1)
dx_1dx_2\cdots dx_N\nonumber\\
=&N\int_0^\infty e^{-r^2}r^{N-1+N(N-1)}dr \int_\Omega
\Delta^2(\Omega)f(r\Omega_1)d\Omega\nonumber\\
=&\int_0^\infty e^{-r^2}r^{N-1+N(N-1)}dr \int_{-1}^1 d\Omega_1
\sigma_{v}(\Omega_1)f(r\Omega_1)
 \nonumber\\
=const.&\int_0^\infty e^{-r^2}r^{N-1+N(N-1)-1}dr \int_{-r}^r dx_1
\sigma_{v}(x_1/r)f(x_1)
 \nonumber\\
=&2\Gamma^{-1}\int_{|x_1|}^\infty e^{-r^2}r^{N-1+N(N-1)-1}dr
\int_{-\infty}^\infty dx_1 \sigma_{v}(x_1/r)f(x_1).\label{3.3}%
\end{align}
 The constant in (\ref{3.3}) is computed by Proposition 1.
\end{proof}

 Similarly, we can derive the integral equation of two point
cluster function of fixed trace square ensemble  $Y_{v,2}(x_1,x_2)$ in terms
of the cluster function of GUE $Y_{GUE,2}(x_1,x_2)=(\frac{\sin{\pi(x_1-x_2)}}{\pi(x_1-x_2)})^2$.

It immediately follows that the asymptotic behavior of $\sigma_{v}(0)$.
\begin{corollary}
\begin{equation}
\sigma_{v,N}(0)=\frac{\sqrt{2N}\Gamma(N^2/2)}{\pi\Gamma(N^2/2-1/2)}(1+o(1)).
\end{equation}
\end{corollary}

\section{main theorem}
 It is well known that the eigenvalue distribution of GUE, $\rho(x)=\sqrt{\frac{2}{N}}\sigma_{GUE,N}(\sqrt{2N}x)$ goes to semicircle:
\begin{equation}
\rho(x)\rightarrow S(x)=\left\{\begin{array}{c}
         \sqrt{1-x^2}, \ if \ |x|\leq 1  \\
           0, \ else. \  \end{array}\right.\label{semicircle}%
\end{equation}
\begin{theorem}
It is possible to prove the asymptotic behavior of the integral equation (\ref{124}). In fact,
The normalized eigenvalue distribution of fixed trace square ensemble $\rho_{v,N}(x)=\frac{2}{N^{\frac{3}{2}}}\sigma_{v,N}(\frac{2x}{\sqrt{N}})$ goes to the semicircle (\ref{semicircle}).
\end{theorem}

\begin{proof}

First we have to estimate the $L^\infty$ norm of $\sigma_N$.

Define a function
\begin{equation}
\xi(x)=\inf\{\xi; \sigma_{v,N}(\frac{x}{\xi})2\Gamma^{-1}(N^2/2)\int_{|x|}^\infty e^{-r^2}r^{N^2-2}dr\}
\end{equation}
Note that $\xi(x)$ is a continuous function on $[0,\infty)$.
Then $\frac{x}{\xi(x)}$(at $x=0$, this function is defined to be zero) is also a continuous function.
Since $\sigma_{GUE,N}\rightarrow 0$(\cite{E}) as $x\rightarrow\infty$,  then

\begin{equation}
\sigma_{v,N}(x/\xi(x))=\frac{\sigma_{GUE,N}(x)}{2\Gamma^{-1}(N^2/2)\int_{|x|}^\infty e^{-r^2}r^{N^2-2}dr}
\rightarrow 0,
\end{equation}
as $x\rightarrow\infty$
Therefore $\sigma_{v,N}(\frac{x}{\xi(x)})\rightarrow 0$ and $\frac{x}{\xi(x)}\rightarrow 1$.
By the expansion of hermite functions in \cite{E}, we know $\forall x\in [0,1]$,
\begin{equation}
\frac{\sigma_{v,N}(x)}{2\Gamma^{-1}(N^2/2)\int_{0}^\infty e^{-r^2}r^{N^2-2}dr}\leq CN^{\frac{1}{2}}.
\end{equation}
Therefore
 \begin{equation}
\sigma_{v,N}\leq CN^{\frac{3}{2}}.
\end{equation}

We shall use the following lemma:
\begin{lemma}$\forall 0<\alpha<1$, $\forall m>0$, if $N$ is sufficiently large,
\begin{equation}
1-N^{-m}\leq \frac{\int_{N/\sqrt{2}-N\alpha}^{N/\sqrt{2}+N\alpha} e^{-r^2}r^{N^2-2}}{\int_0^\infty  e^{-r^2}r^{N^2-2}}dr \leq 1.
\end{equation}
\end{lemma}
The proof is straightforward.

For $\forall \alpha>0$, if $N$ is sufficiently large,

\begin{align}
& 2\Gamma^{-1}(N^2/2)\int_{|x|}e^{-r^2}r^{N^2-2}\sigma_{v,N}(x/r)dr\nonumber\\
=& 2\Gamma^{-1}(N^2/2)(\int_{|x|}^{N/\sqrt{2}-N\alpha}+\int_{N/\sqrt{2}-N\alpha}^{N/\sqrt{2}+N\alpha}+\int_{N/\sqrt{2}+N\alpha}^\infty)e^{-r^2}r^{N^2-2}\sigma_{v,N}(x/r)dr\nonumber\\
=& 2\Gamma^{-1}(N^2/2)(\int_{|x|}^{N/\sqrt{2}-N\alpha}+\int_{N/\sqrt{2}+N\alpha}^\infty)e^{-r^2}r^{N^2-2}N^{1.5}dr+\int_{N/\sqrt{2}-N\alpha}^{N/\sqrt{2}+N\alpha}e^{-r^2}r^{N^2-2}\sigma_{v,N}(x/r)dr\nonumber\\
= & 2\Gamma^{-1}(N^2/2)\int_{N/\sqrt{2}-N\alpha}^{N/\sqrt{2}+N\alpha}e^{-r^2}r^{N^2-2}\sigma_{v,N}(x/r)dr+O(1/N^5)\nonumber\\
= & 2\Gamma^{-1}(N^2/2)\sigma_{v,N}(x/r(x))\int_{N/\sqrt{2}-N\alpha}^{N/\sqrt{2}+N\alpha}e^{-r^2}r^{N^2-2}dr+O(1/N^5),
\end{align}

where $N/\sqrt{2}-N\alpha \leq r(x)\leq N/\sqrt{2}+N\alpha$.
Therefore $\forall 0\leq y$,

\begin{equation}
\sigma_{v,N}(y)=\frac{\Gamma(N^2/2)}{2\int_{N/\sqrt{2}-N\alpha}^{N/\sqrt{2}+N\alpha}e^{-r^2}r^{N^2-2}dr}\sigma_{GUE,N}(r(\tilde{y})y)+O(1/N^4).
\end{equation}

We notice  that $\frac{\Gamma(N^2/2)}{2\int_0^\infty e^{-r^2}r^{N^2-2}dr}\rightarrow N/\sqrt{2}+O(1/\sqrt{N})$(see \cite{F}).
Therefore,

\begin{align}
&|\rho_{v,N}(y)-S(y)|\nonumber\\
\leq & |\frac{2}{N^{\frac{3}{2}}}\sigma_{v,N}(\frac{2y}{\sqrt{N}})-S(y)|\nonumber\\
\leq &|\sqrt{\frac{2}{N}}\sigma_{GUE,N}(\tilde{r}2y/\sqrt{N})- S(y)|+ o(1)\nonumber\\
\leq &|\sqrt{\frac{2}{N}}\sigma_{GUE,N}(\sqrt{2N}(\tilde{r}2y/(\sqrt{N}\sqrt{2N})))-S(\tilde{r}2y/(\sqrt{N}\sqrt{2N}))|+
|S(\tilde{r}2y/(\sqrt{N}\sqrt{2N}))- S(y)|+ o(1)\nonumber\\
\leq & \sup_{|x-y|\leq \alpha} |S(x)-S(y)|+ o(1)
\end{align}
\end{proof}

 We use the software Matlab to estimate the level density $\sigma_{\Phi,N}$
of fixed trace square ensemble in the case $N=10$, $N=50$, and $N=100$.
Surprisingly, these graphs look really like semicircles.

\begin{figure}
\centering
\includegraphics[height=10cm]{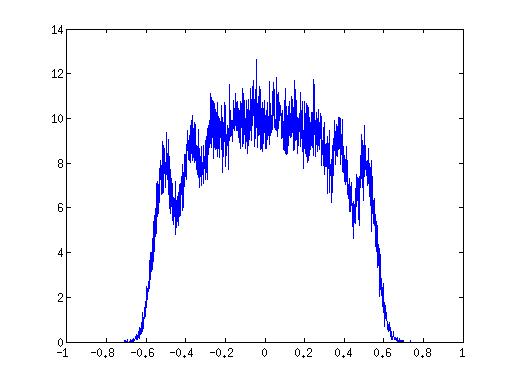}
\caption{Level density of fixed trace square ensemble N=10} \label{fig1}

\includegraphics[height=10cm]{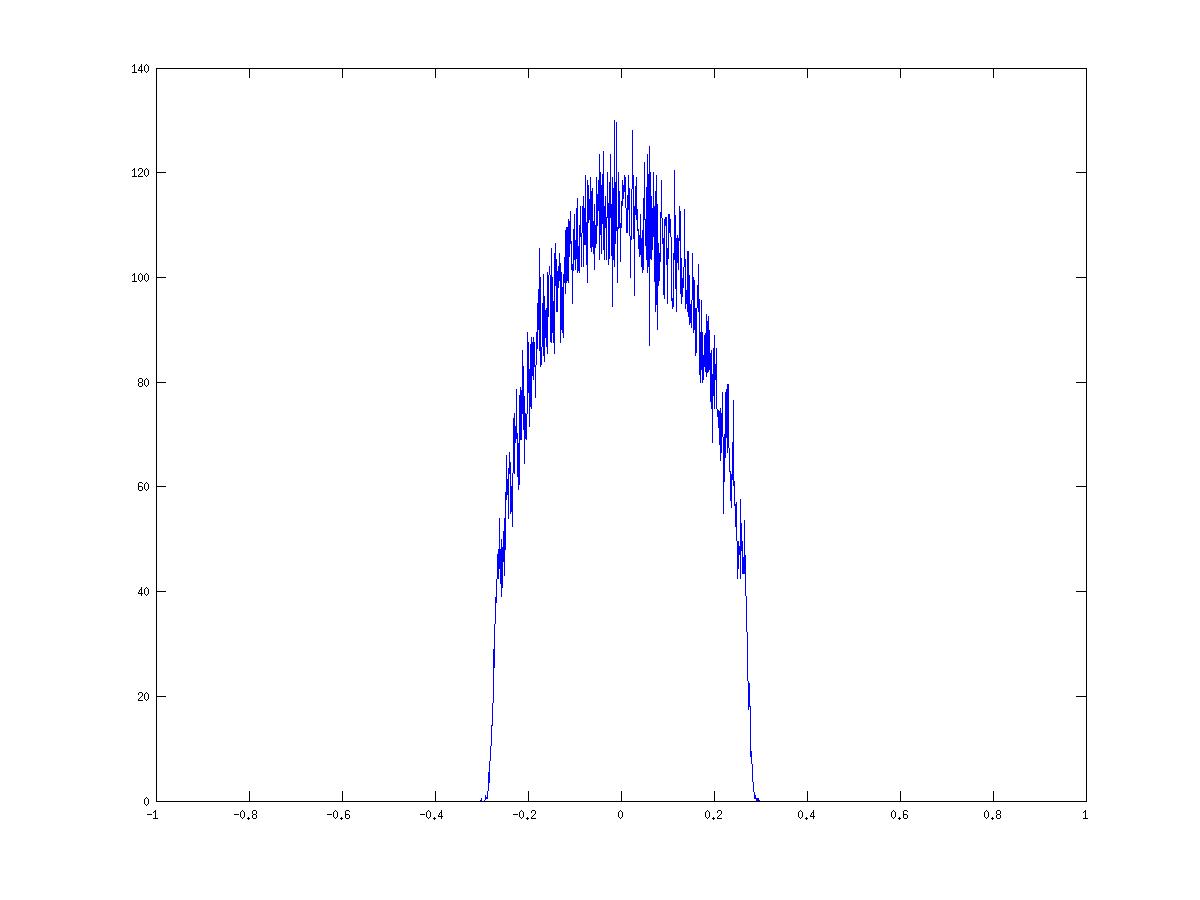}
\caption{Level density of fixed trace square ensemble N=50} \label{fig2}
 \end{figure}

 \begin{figure}
 \centering
\includegraphics[height=10cm]{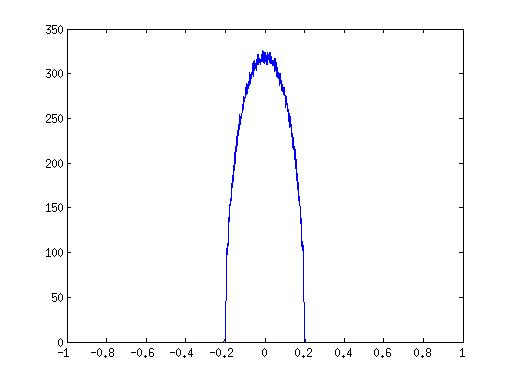}
\caption{Level density of fixed trace square ensemble N=100} \label{fig3}

\includegraphics[height=10cm]{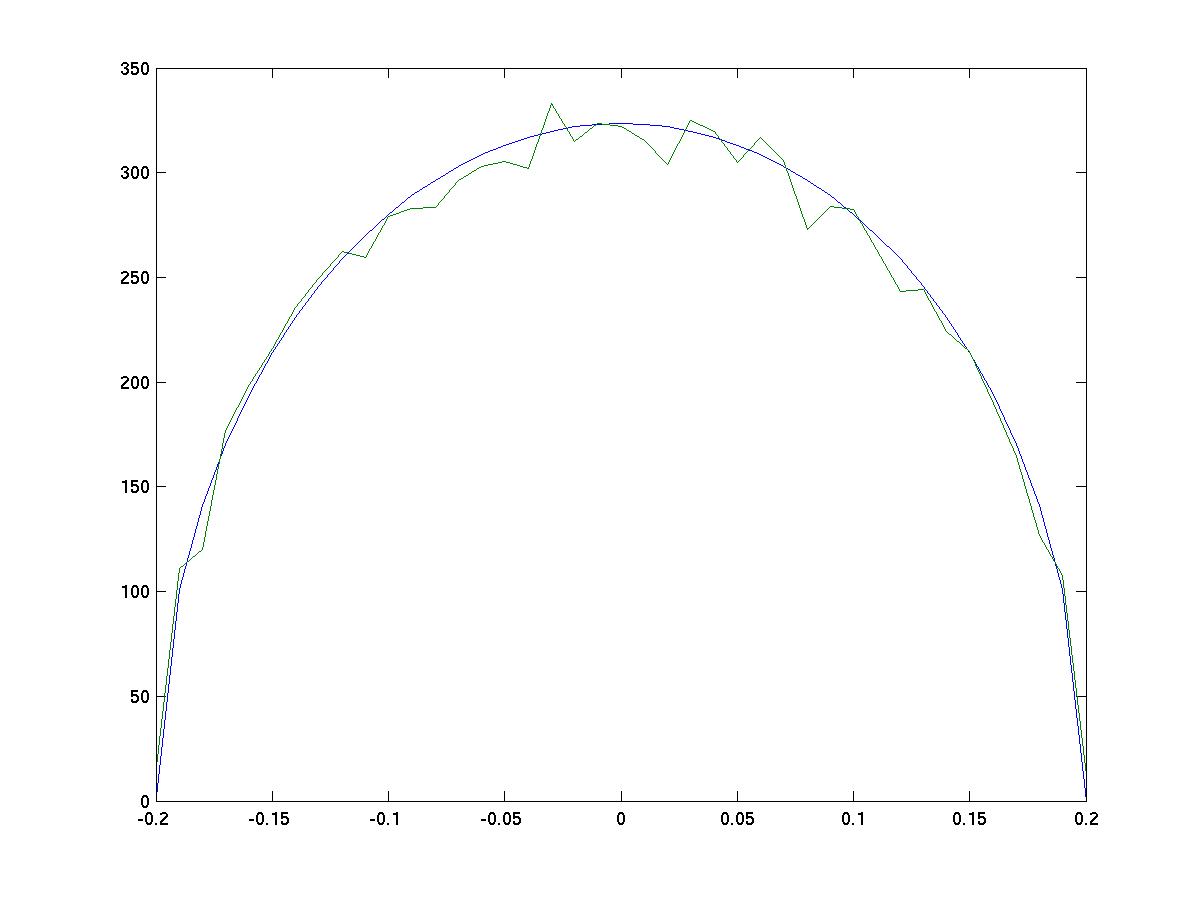}
\caption{the coincidence of the semicircle and the level density when $N=100$} \label{fig4}
\end{figure}

\end{document}